\documentclass{PoS}
\usepackage{amsmath}
\usepackage{subcaption}

\newcommand{\LHAPDF}{\textsc{LHAPDF}~}

\title{Efficient interpolation and evolution of parton distribution functions}

\ShortTitle{Efficient interpolation and evolution of parton distribution functions}

\author{\speaker{Riccardo Nagar}\thanks{The work presented here is part of a collaboration with Markus Diehl and Frank Tackmann (DESY).}\\
        Deutsches Elektronen-Synchrotron DESY\\
        Notkestra\ss e 85, D-22607 Hamburg, Germany\\
        E-mail: \email{riccardo.nagar@desy.de}}

\abstract{We present an efficient numerical solution of the DGLAP equations for single and double parton distribution functions (PDFs and DPDs), based on the Chebyshev interpolation of these functions.
For PDF evolution, our method allows for a higher numerical accuracy using a considerably smaller number of grid points compared to other methods. The DPD evolution is realized using an affordable number of grid points, and allows for two independent renormalization scales for the two partons. Both methods include NNLO DGLAP kernels and flavor matching.}

\FullConference{XXVII International Workshop on Deep-Inelastic Scattering and Related Subjects - DIS2019\\
		8-12 April, 2019\\
		Torino, Italy}

\graphicspath{{plots/}}

\begin{document}

\paragraph{Introduction}

Parton distribution functions (PDFs) are an essential ingredient
of virtually all phenomenology involving a hadron in the initial state.
The shape of these distributions can be fitted from experimental data,
and their value at any arbitrary energy scale can be computed by solving the DGLAP evolution equations.
One approach to solve this system of integro-differential equations
consists in discretizing the PDF set $f_a(x, \mu)$ on a grid in $x$,
and is currently implemented in several publicly-available packages
\cite{Bertone:2013vaa, Salam:2008qg, Botje:2010ay}.
The interface \LHAPDF \cite{Buckley:2014ana} performs a fast interpolation on similar grids in $x$ and $\mu$,
and provides a unified and easy access to all the main PDF sets on the market.

Recent developments in precision calculations highlight a few cases
in which the aforementioned algorithms might be inadequate.
As it has been shown for example in \cite{Dulat:2017prg},
the numerical accuracy of the PDF interpolation algorithms can be insufficient in higher-order calculations,
which usually involve computing Mellin convolutions with very singular distributions.

On another note, the double parton scattering (DPS) cross-section formula
is characterized by the presence of double parton distributions (DPDs).
These distributions $F_{a_1 a_2} (x_1, x_2, y, \mu_1, \mu_2)$ evolve according to generalized DGLAP equations \cite{Diehl:2011yj}.
Since each DPD depends on many variables,
the discretization of a single DPD set on grids similar to the ones used for PDFs involves a large amount of data,
making a straightforward extension of the current evolution methods computationally unfeasible or very hard.
Numerical solutions have been developed in \cite{Gaunt:2009re, Elias:2017flu}.

We present a different approach to the approximate solution of the DGLAP equations,
based on the Chebyshev interpolation of PDFs.
With our approach we can achieve a numerical accuracy
which is orders of magnitude higher than the typical one,
using considerably smaller grids.
The efficiency of this method applied to PDF evolution
opens the way to its full extension to the generic DPD case,
i.e.~with dependence on the two renormalization scales $\mu_1$ and $\mu_2$,
and on the transverse separation $y$.
This algorithm is being implemented in a \textsc{C++} library
called \textsc{ChiliPDF} (Chebyshev interpolation library for PDFs),
to be made publicly available.
\textsc{ChiliPDF} can perform DGLAP evolution up to NNLO, and flavor matching up to $O(\alpha_s^2)$
with free choice of matching scales.
It accepts any user-given set of starting PDFs or DPDs.
The code development strategy is based on modular design,
making the library readily extensible to diversified use-cases.
More detail will be given in \cite{Diehl:2019uuu}.

\paragraph{The interpolation algorithm}

The numerical methods commonly used to interpolate PDFs are based on the discretization of a PDF $f_a(x, \mu)$
on one or more grids equispaced in a transformed variable $u$.
In most cases $u = \log x$, however some grids have a more complex variable transformation,
and at high-$x$ it is common to use directly the grid variable $u = x$.
We denote the grid points as $x_k$,
the corresponding transformed grid points as $u_k$,
and the values of the function in the grid points as $f_k \equiv f(x_k)$.
The common aspect of these algorithms
is the constant separation between the consecutive $u_k$'s.
The function that is actually handled by these algorithms is $\tilde{f}(x) = x f(x)$,
and we make the same choice for our method.

The interpolation routines corresponding to these equispaced grids
involve polynomials of low degree (usually not over five),
and implement either splines or Lagrange polynomials
on a small subset of the $x$-grid surrounding the interpolation point.

For our method, we pick the same variable transformation $u = \log x$,
but for the $\{ u_k \}$ grid-points we take instead the Chebyshev points
mapped onto the interval $[ u_\text{min}, u_\text{max} ]$ via a linear transformation
on the unshifted Chebyshev points $\xi_k = \cos ( k \pi / N )$,
defined on the interval $[-1, 1]$.
$N$ is the Chebyshev polynomial degree.
There are $N+1$ points in a $N$th degree grid.

The interpolated value $\tilde{f}(x)$ is obtained given the vector $\tilde{f}_k$
using the \emph{barycentric formula} \cite{Trefethen:2013}
\begin{equation}
   \tilde{f}(x) = \sum_{j = 1}^N \tilde{f}_j \, b_j(u) \vert_{u = \log x} \qquad
   \text{ with } \quad
   b_j(u) = \dfrac{(-1)^j \beta_j}{u - u_j} \bigg/ \displaystyle \sum_{i=0}^N \dfrac{(-1)^i \beta_i}{u - u_i}\, ,
   \label{eqn:barycentric}
\end{equation}
where $\beta_0 = \beta_N = 1/2$ and $\beta_i = 1$ otherwise.
This formula evaluates the $N$th order polynomial
passing through all the points $(x_k, f_k)$
with a computational complexity of $O(N)$.

\begin{figure}
   \centering
   \begin{subfigure}[b]{0.49\textwidth}
   \includegraphics[width=\textwidth]{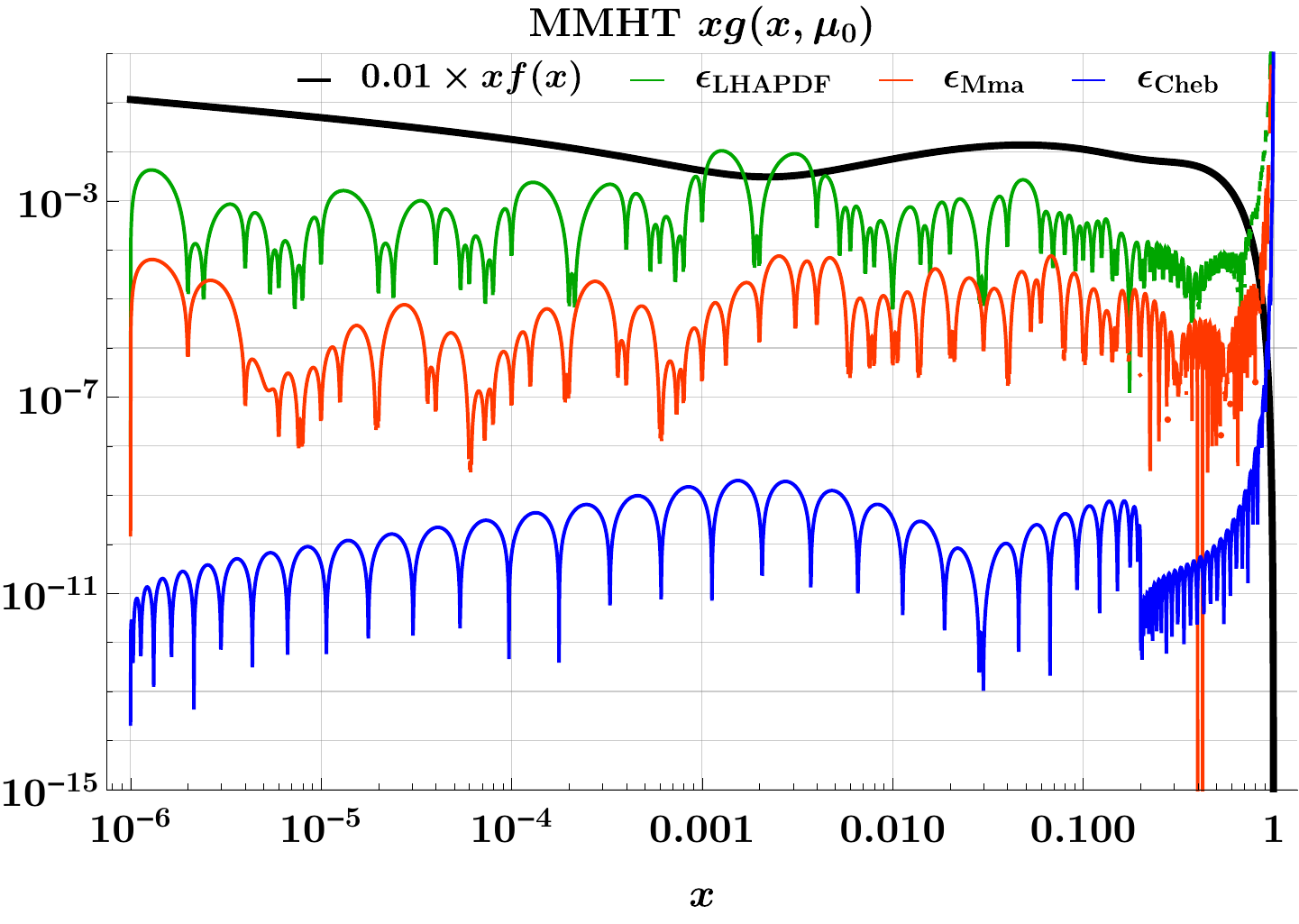}
   \end{subfigure}
   \,
   \begin{subfigure}[b]{0.49\textwidth}
   \includegraphics[width=\textwidth]{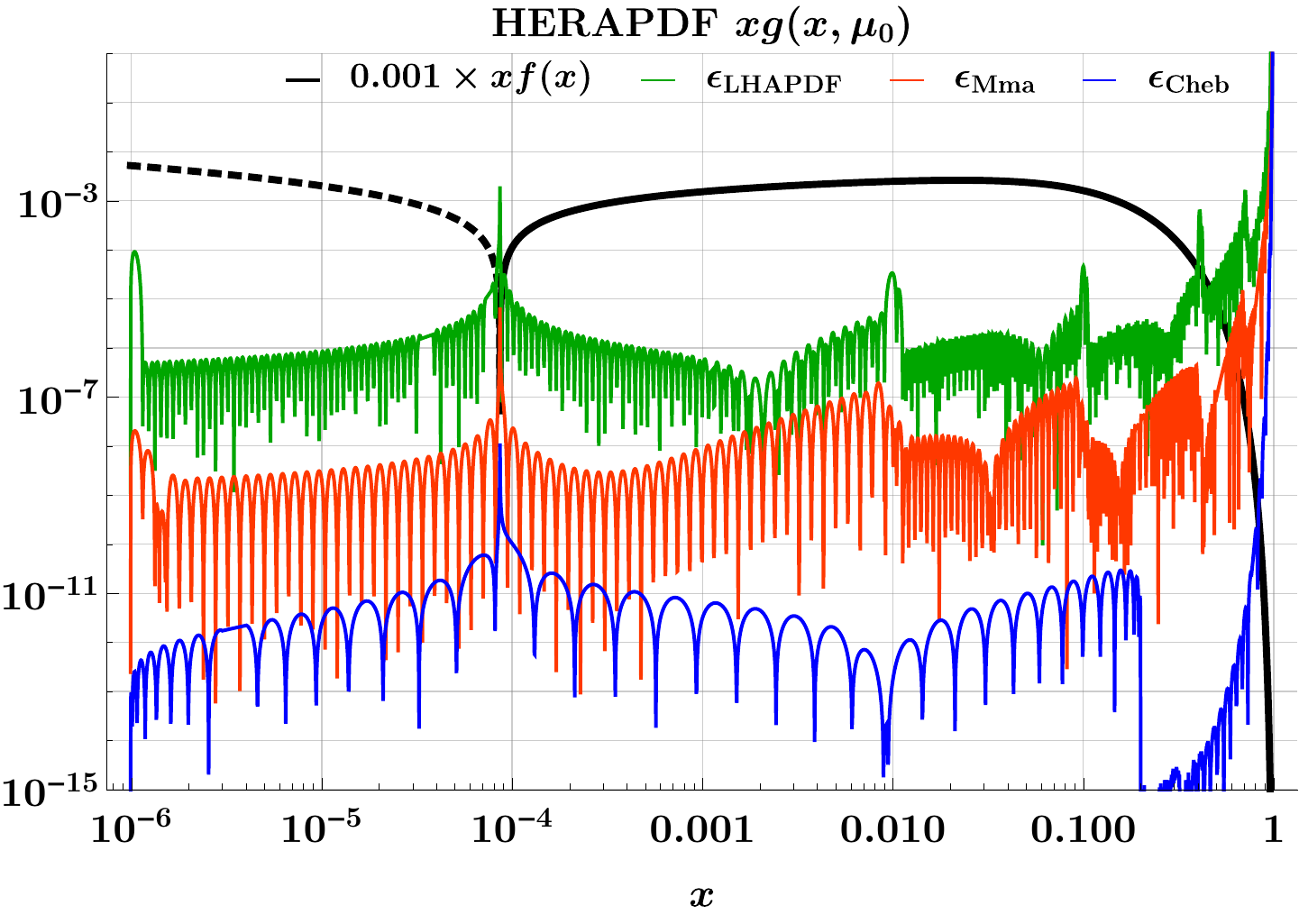}
   \end{subfigure}
   \caption{
   Comparison of the Chebyshev interpolation relative accuracy (in blue)
   with respect to the accuracy obtained using two different splines methods:
   the one corresponding to the LogBicubic interpolator of \textsc{LHAPDF} 6.2 (in green),
   and the one corresponding to the default Mathematica 11 interpolator (in red).
   The interpolated function is shown in black,
   and is dashed where its value is negative.
   }
   \label{fig:interpolation_comparison}
\end{figure}

The efficiency of this method with respect to two different example grids is shown in Figure~\ref{fig:interpolation_comparison}.
On the left the MMHT starting-scale gluon PDF \cite{Harland-Lang:2014zoa} is plotted,
using the default MMHT 64-points grid for the splines interpolations,
and a Chebyshev grid composed of two sub-grids $[10^{-6}, 0.1]$ and $[0.1, 1]$ with a total of 63 points for our interpolation.
The plot on the right shows the HERAPDF starting-scale gluon PDF \cite{Zhang:2015tuh},
using the default HERAPDF 199-points grid for the splines, and a composite Chebyshev grid with a total of 71 points.

The plots show that the Chebyshev interpolation method can reach relative accuracies
which are orders of magnitude higher than the typical splines accuracy
even with a limited number of grid points.
We verified that the same happens for different PDF functional forms,
including PDFs at higher scales.
We also checked against additional grids like the ones used by other PDF fitting groups.

\paragraph{Mellin convolution}

The discretization
simplifies the Mellin convolution of a PDF with an arbitrary kernel.
Using the barycentric formula \eqref{eqn:barycentric}, one can evaluate the convolution $(K \otimes f) (x)$ in the grid points as
\begin{equation}
   (K \otimes f)_m = (K \otimes f) (x_m) = K_{mn} \, f_n,
   \qquad \text{with } \quad
   K_{mn} = \int_{x_m}^1 \! \frac{dz}{z} \, K(z) \, \hat{b}_n \left( \frac{x_m}{z} \right) \, ,
   \label{eqn:mellin_discretized}
\end{equation}
where $\hat{b}(x) = b(u) \vert_{u = \log x}$.
One can obtain the convolution at any other value of $x$ via barycentric interpolation
using the $(K \otimes f)_m$ as the sampled function values.
The matrix $K_{mn}$ does not depend on the function $f(x)$,
hence it can be pre-computed and used for the convolution with any function.

\begin{figure}
   \centering
   \begin{subfigure}[b]{0.49\textwidth}
   \includegraphics[width=\textwidth]{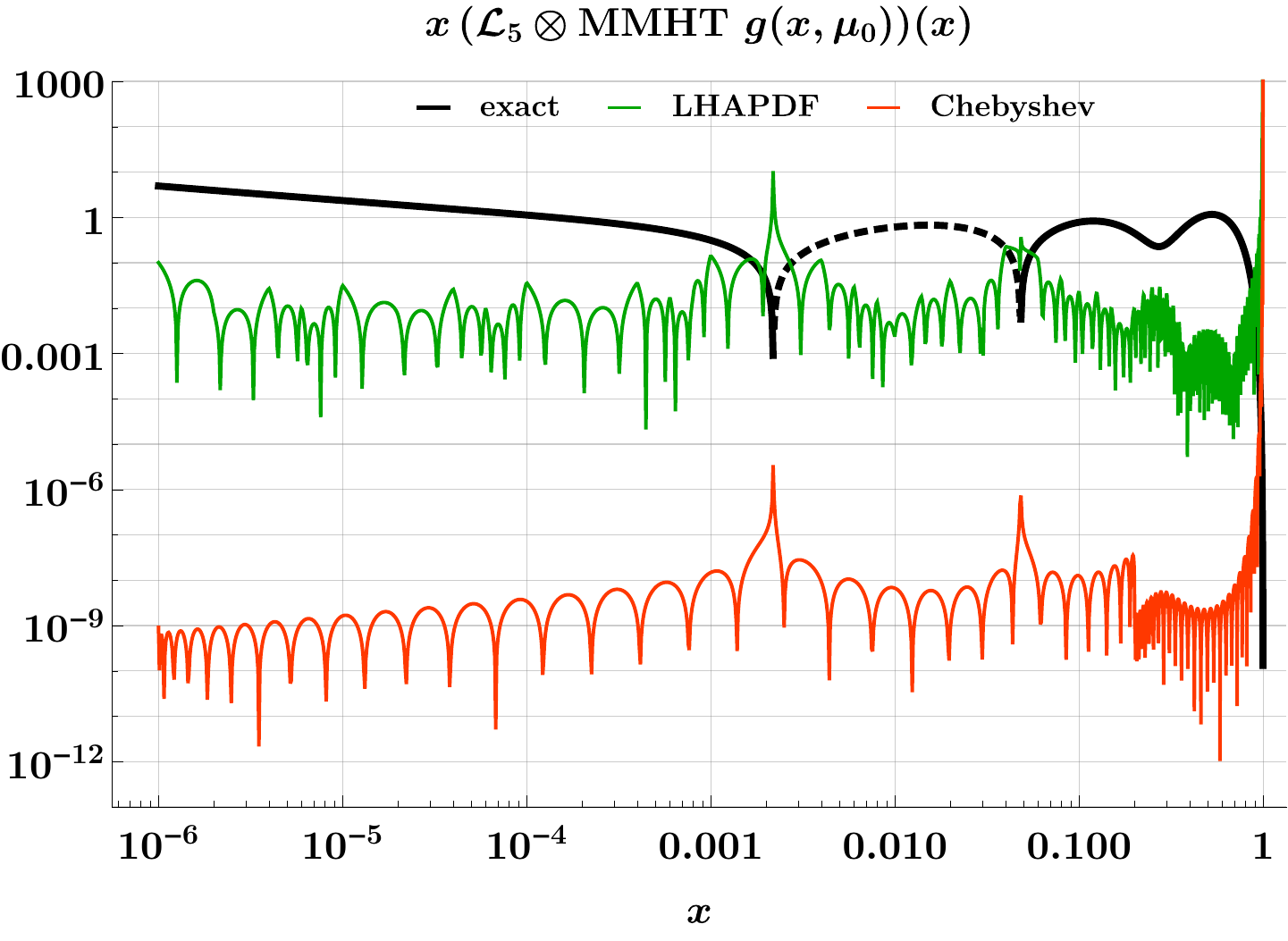}
   \caption{$x \, (\mathcal{L}_5 \otimes g) (x)$}
   \label{fig:mellin_comparison_L5}
   \end{subfigure}
   \,
   \begin{subfigure}[b]{0.49\textwidth}
   \includegraphics[width=\textwidth]{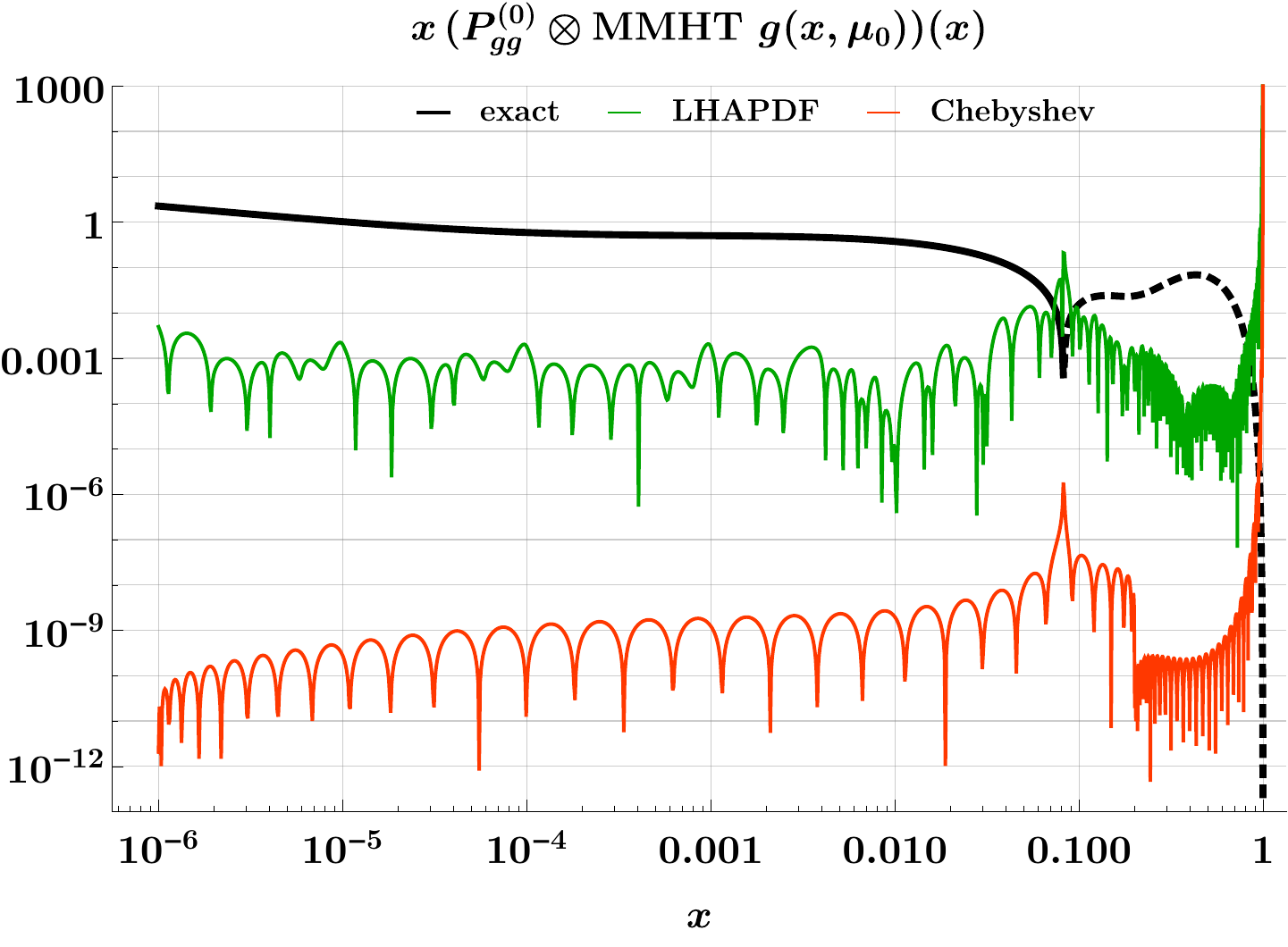}
   \caption{$x \, (P_{gg}^{(0)} \otimes g) (x)$}
   \label{fig:mellin_comparison_Pgg0}
   \end{subfigure}
   \caption{
   Comparison of the relative accuracy of the Mellin convolution
   calculated using the direct \LHAPDF input (in green)
   and using our method (in red).
   The \LHAPDF input is based on the MMHT $x$-grid of 64 points.
   The Chebyshev implementation uses a grid with a total of 63 points.
   The exact result of $x \, (K \otimes f) (x)$ is shown in black,
   and is dashed where its value is negative.
   }
   \label{fig:mellin_comparison}
\end{figure}

We compared the convolution calculated with our method
against the one obtained by explicitly performing the convolution integral
using the out-of-the-box \LHAPDF call for the PDF part of the integrand.
We analyzed various typical kernels, PDF functional forms, and \LHAPDF grids.
Two examples are shown in Figure~\ref{fig:mellin_comparison},
namely the high-order plus distribution
$\mathcal{L}_5 (z) = [ \log^5 (1-z) / (1-z) ]_{+}$ in Figure~\ref{fig:mellin_comparison_L5},
and the splitting function $P_{gg}^{(0)} (z)$ at $O(\alpha_s)$ in Figure~\ref{fig:mellin_comparison_Pgg0}.
Both kernels are convolved with the same distribution shown in the left-hand side of Figure~\ref{fig:interpolation_comparison}.

\paragraph{DGLAP evolution}

The DGLAP equations are solved in $x$-space by discretizing the PDFs and the Mellin convolutions,
therefore obtaining the linear system of ODEs
\begin{equation}
   \text{d} \tilde{f}^a_m \, / \, \text{d} \log \mu = \tilde{P}^{ab}_{mn} \, \tilde{f}^b_n \, ,
   \label{eqn:dglap_discretized}
\end{equation}
where $\tilde{P}^{ab}_{mn}$ is the analog of $K_{mn}$ in equation \eqref{eqn:mellin_discretized}
for the function $z \, P^{ab}(z)$.
The singlet and gluon PDFs evolve according to a coupled set of equations.

DPDs obey a double set of DGLAP equations,
one with a convolution in $x_1$ and another in $x_2$
\begin{equation}
   \text{d} F_{a_1 a_2} (x_1, x_2, y, \mu_1, \mu_2) \, / \text{d} \log \mu_i =
   ( P^{(i)} \underset{i}{\otimes} F_{a_1 a_2} ) (x_1, x_2, y, \mu_1, \mu_2) \, ,
\end{equation}
where the $i$ below the convolution symbol denotes the index of the convolution variable.
These equations can be discretized like in \eqref{eqn:dglap_discretized}.
The two equations can be decoupled,
allowing one to use the same matrices $\tilde{P}^{ab}_{mn}$
computed for PDF evolution.
Evolving from $(\mu^{(0)}_1, \mu^{(0)}_2)$ to any pair of scales $(\mu_1, \mu_2)$
consists in fact of separate and interchangeable evolution steps $\mu^{(0)}_1 \to \mu_1$ and $\mu^{(0)}_2 \to \mu_2$.

The system of differential equations is solved numerically
by implementing a Runge-Kutta (RK) algorithm.
Given an evolution step $h$,
a RK algorithm comes with a local relative error of $O(h^p)$,
where $p$ is a positive integer.
We adopted different versions of the RK algorithm:
the ``classic'' RK of order $h^4$,
the Cash-Karp method of order $h^5$,
and the Dormand-Prince (DOPRI) method of order $h^6$.
The RK accuracy can be optimized by varying the method and the step size.

We compared the results of our algorithm with the benchmarks established in
\cite{Giele:2002hx, Dittmar:2005ed}.
We found agreement with the benchmark evolution,
using a composite Chebyshev grid on the interval $[10^{-8}, 1]$
composed of three sub-intervals ($[10^{-8}, 10^{-3}]$, $[10^{-3}, 0.5]$, and $[0.5, 1]$),
with a total of 69 grid points, and a DOPRI RK with step-size $h = 0.02$.
In comparison, \textsc{hoppet} obtains the benchmark results
using a total of 1,170 points in a composite grid in $x \in [10^{-8}, 1]$
and 220 points in $\mu^2 \in [2, 10^6] \text{ GeV}^2$.

\paragraph{Results}

We study the accuracy reach of our algorithm
by comparing results obtained with $x$-grids of different densities and RK methods with different step-sizes.

\begin{figure}
   \centering
   \begin{subfigure}[b]{0.49\textwidth}
   \includegraphics[width=\textwidth]{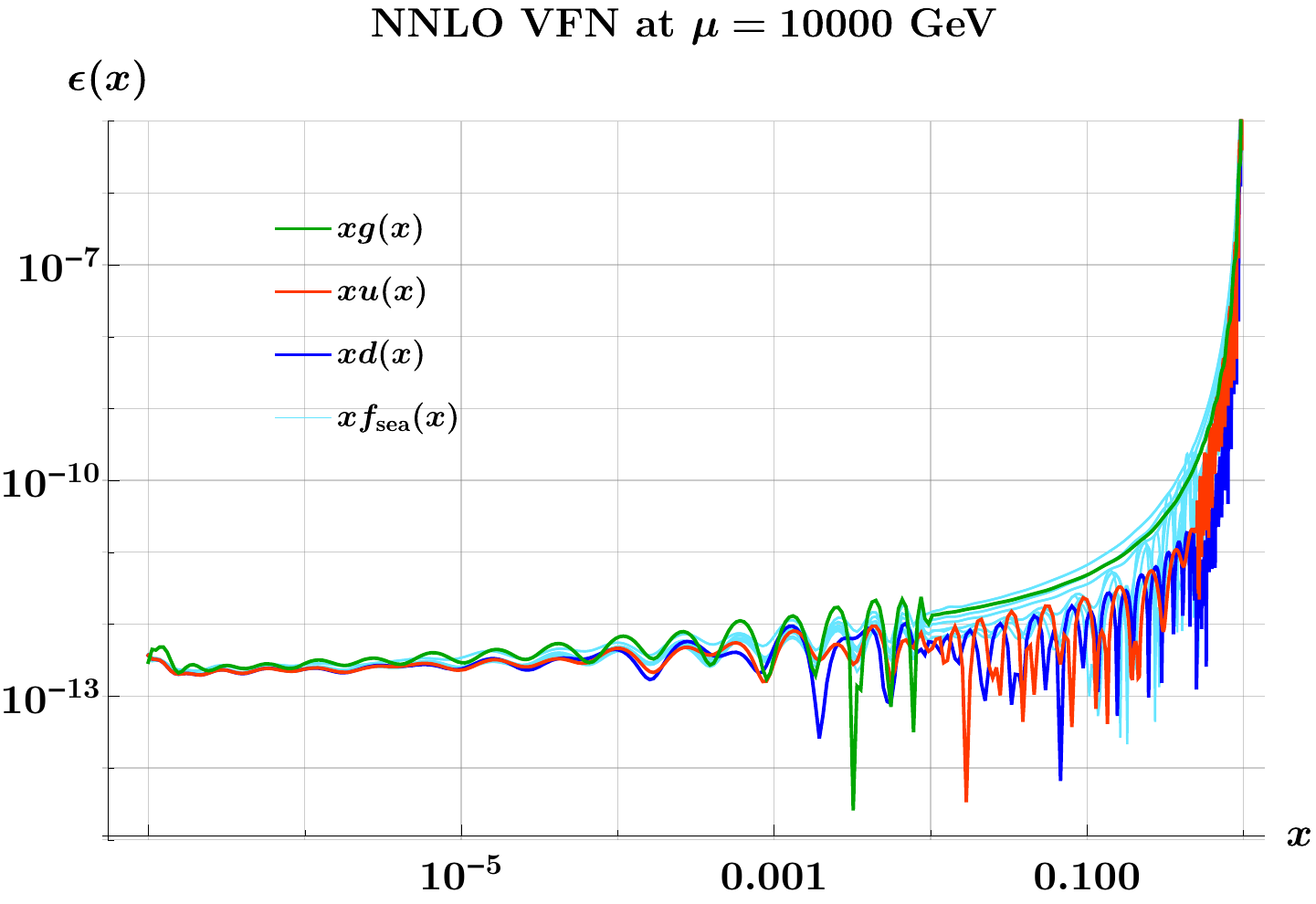}
   \caption{Quarks, antiquarks, and gluon.}
   \label{fig:pdf_accuracy_all}
   \end{subfigure}
   \,
   \begin{subfigure}[b]{0.49\textwidth}
   \includegraphics[width=\textwidth]{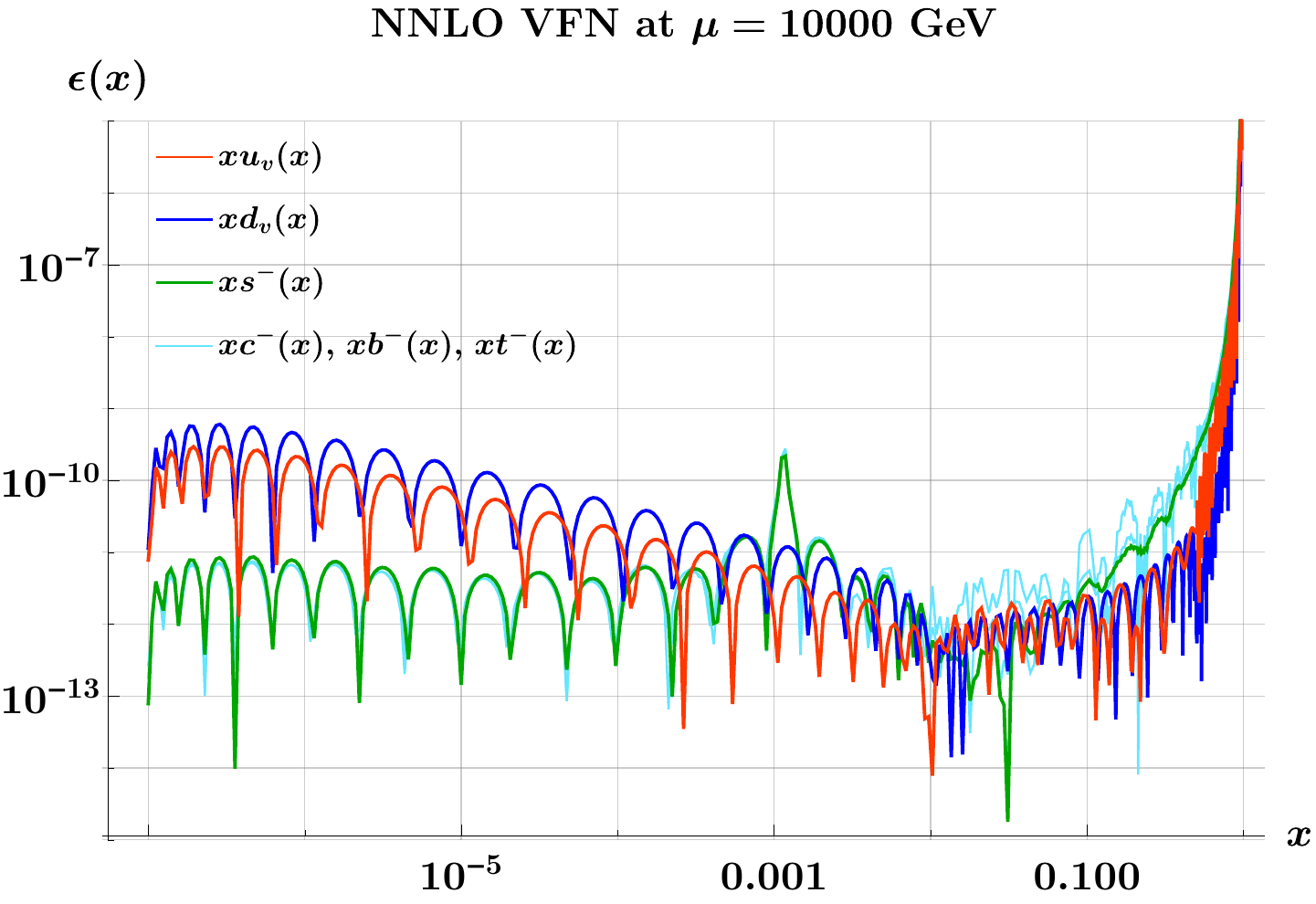}
   \caption{Valence distributions.}
   \label{fig:pdf_accuracy_val}
   \end{subfigure}
   \caption{
   Relative accuracies in the evolution of a full PDF set.
   The starting PDF set parametrization is taken from \cite{Giele:2002hx}.
   }
   \label{fig:pdf_accuracy}
\end{figure}

For PDF evolution and flavor matching the results are shown in Figure~\ref{fig:pdf_accuracy}.
In this study we take $x \in [ 10^{-7}, 1 ]$ and evolve up to $\mu = 10 \text{ TeV}$ using the DOPRI RK method.
We compare the results obtained with a grid of 71 points and $h = 0.02$
with the ones obtained with a grid of 107 points and $h = 0.004$.
The relative accuracy is better than $10^{-8}$ for $x < 0.8$,
and orders of magnitude lower in the small-$x$ region.
By a differential analysis we determined that, given our settings,
the error relative to the RK step size is always negligible
with respect to the error introduced by the discretization in $x$.

\begin{figure}
   \centering
   \begin{subfigure}[b]{0.49\textwidth}
   \includegraphics[width=\textwidth]{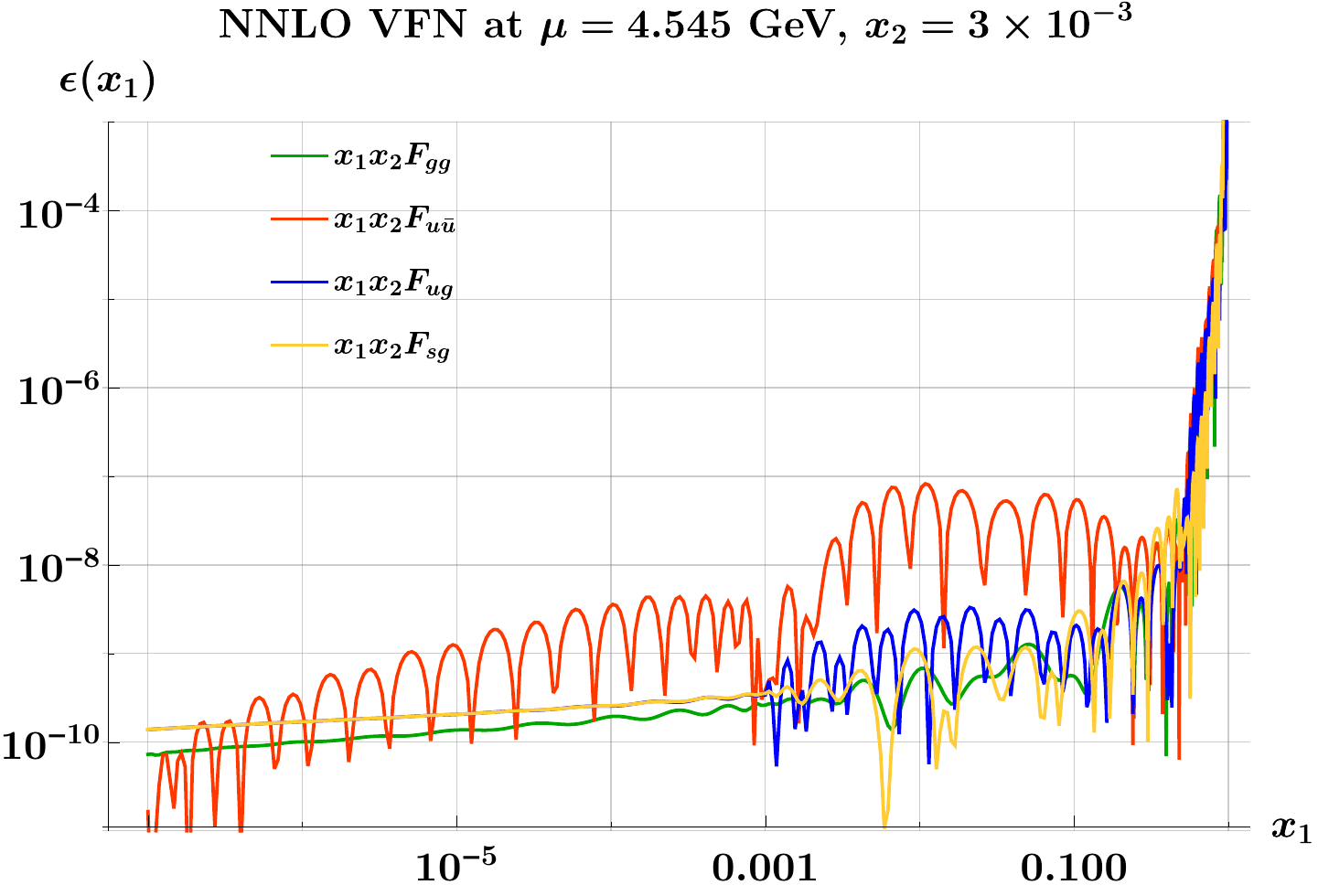}
   \end{subfigure}
   \,
   \begin{subfigure}[b]{0.49\textwidth}
   \includegraphics[width=\textwidth]{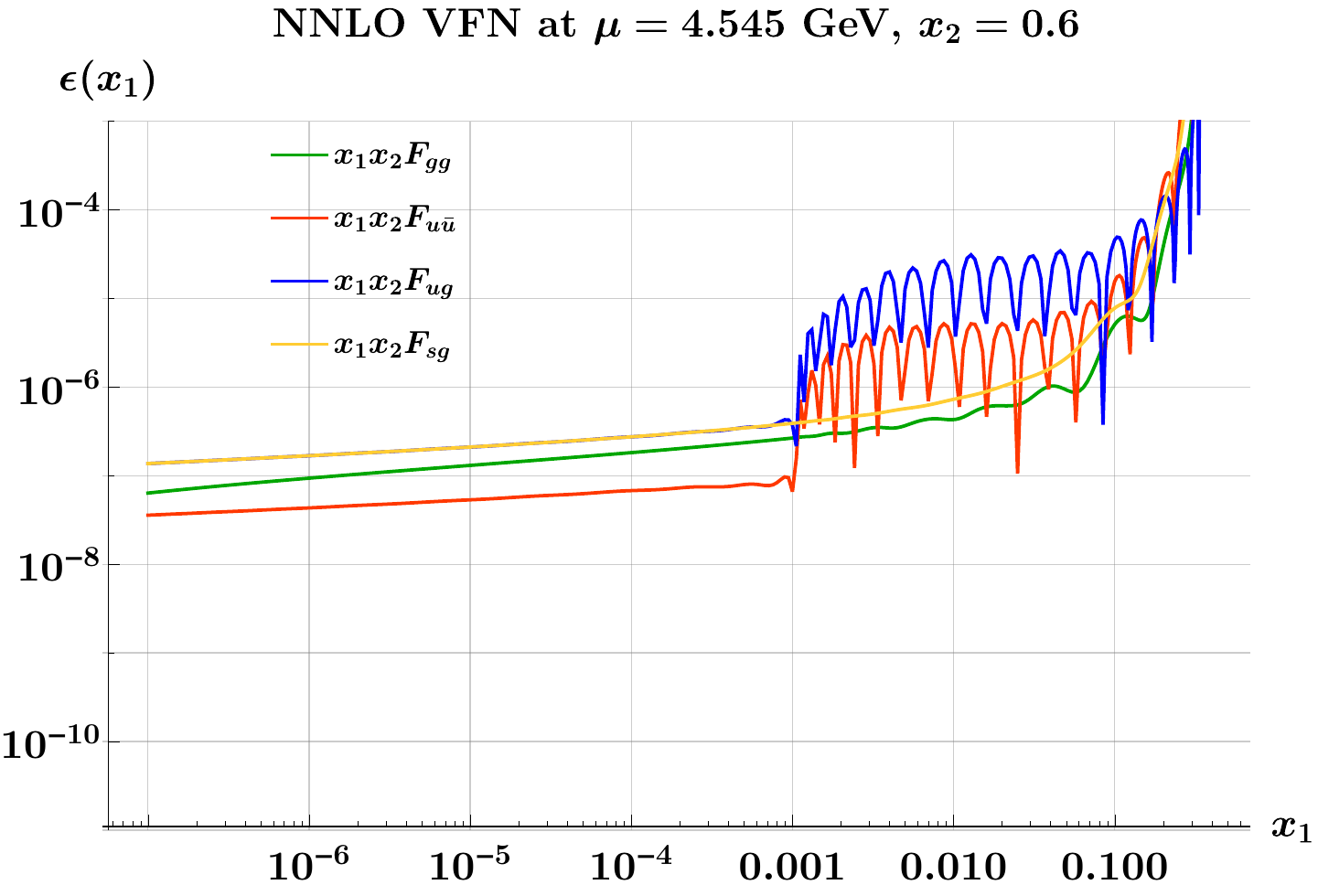}
   \end{subfigure}
   \caption{
   Relative accuracies in the evolution of a full DPD set.
   The plots show a representative sample of four DPDs ($F_{gg}$, $F_{u\bar{u}}$, $F_{ug}$, and $F_{sg}$) at two fixed values of $x_2$ ($3 \times 10^{-3}$ and 0.6).
   }
   \label{fig:dpd_accuracy}
\end{figure}

For DPD evolution and flavor matching the resulting accuracy is shown in Figure~\ref{fig:dpd_accuracy}.
Here we take the same grids as in the PDF evolution accuracy study, but we compare the cases $h = 0.05$ and $h = 0.01$.
In analogy to the PDF case, the degradation in the accuracy occurs
in the region close to the kinematical limit $x_1 + x_2 = 1$.
This is most evident in the plot at the right-hand side of Figure~\ref{fig:dpd_accuracy}.
Overall we get an accuracy better than $O(10^{-4})$ in the region where $x_1 + x_2 < 0.8$.

\bibliographystyle{JHEP}
\bibliography{DIS2019proc}

\end{document}